\documentclass[preprint]{aa}

\usepackage{graphicx}
\usepackage{natbib} 
\newcommand{\be}{\begin{eqnarray}}
\newcommand{\ee}{\end{eqnarray}}

\newcommand {\nbodypp}{\textsc{\mbox{nbody6\raise.4ex\hbox{\tiny++}}}}
\newcommand {\COri} {\mbox{$\theta^1{\rm{C}}\:{\rm{Ori}}$}}

\newcommand {\Msun} {\mbox{M$_{\odot}$}}

\begin{document}

\title{Quasi-binarity of massive stars in young dense clusters - the case of the ONC}
\author{S.~Pfalzner \and C.~Olczak} 
\institute{I. Physikalisches Institut, University of Cologne, Z\"ulpicher Str. 77, 50937 Cologne, Germany}
\authorrunning{Pfalzner \& Olczak}
\titlerunning{Quasi-binarity of massive stars}

\abstract
{Observations indicate that in young stellar clusters the binary fraction for 
massive stars is higher than for solar mass stars. For the Orion Nebula Cluster
(ONC) there is a binary frequency of $\sim$ 50\% for solar-mass stars 
compared to 70-100\% for the massive O- and B-stars.}
{We explore the reasons for this discrepancy and come up with two possible answers: 
a) a primordially higher binarity of massive stars could be inherent to the star 
formation process or b) the primordial binary rate might be the same for solar-mass 
and massive stars, but the higher capture cross section of the massive stars possibly 
leads to the formation of additional massive binaries in the early cluster development. Here we 
investigate the likelihood of the latter scenario in detail using the ONC as an example.}
{N-body simulations are performed to track the capture events in an ONC-like cluster.} 
{We find that whereas low-mass stars rarely form bound systems through capture, the dynamics 
of the massive stars - especially in the first 0.5 Myrs - is dominated by a rapid 
succession of  ``transient binary or multiple 
systems''.
In observations the transient nature of these systems would not be apparent, so
that they would be rated as binaries.  At 1-2 Myrs, the supposed age of the ONC, the 
``transient'' massive systems become increasingly stable, 
lasting on average several 10$^6$ yrs. Despite the ONC being so young, the observed binary 
frequency  for massive stars --- unlike that of solar-mass stars --- is not identical to the 
primordial binary frequency but is increased by at least 10-15\% through dynamical interaction 
processes. This value might be increased to at least 20-25\% by taking disc effects into account.} 
{The primordial binary frequency 
could well be the same for massive and solar mass stars because the observed difference
can be explained by capture processes alone.}


\maketitle

\section{Introduction}

Observations predominantly in the 1990's found the fraction of field stars in multiple systems to 
be high --- for solar-mass stars $\sim$ 55$\%$, \cite{duquennoy:91} and for M-dwarf low-mass stars 
$\sim$ 35-42$\%$ \cite{fischer:92,reid:97}.
To explain the high binary rate of field stars, gravitational capture between two single stars 
was initially discussed as
a possibility. However, it was acknowledged that while some binaries might actually
form this way, capture cannot be the primary binary formation process due to the low likelihood 
of such an event. 

Observations of pre-main sequence stars support the theory that binaries 
are already formed during star formation. The multiplicity of young pre-main-sequence (PMS) stars 
in many regions (especially in loose T-associations like Taurus) seems to be systematically higher than that 
of their main-sequence counterparts \citep[e.g.][and references therein]{mat:00}.
Currently fragmentation of the molecular cloud during the formation of protostars is the accept 
explanation for the formation of a binary or multiple star systems \cite{boss:92,tohline}

However, unsolved problems remain. Denser young stellar clusters like the central (very dense) Trapezium 
cluster of the Orion Nebula Cluster, IC 348 and NGC 2024 show much lower binary frequencies 
than above mentioned loose T-associations 
\cite{prosser:94,padgett:97,bouvier:97,patience:98,petr:98,simon:99,scally:99,duchene:99b,mccaughrean:01,beck:03,liu:03,luhman:05,kraus:07}. 
There is emerging evidence that high-density regions might have lower binary frequencies 
or preferentially smaller binary separations \cite{koehler:06}. Essentially two theoretical 
concepts exist to explain this observational finding: (a) the dynamical 
disruption of wide binaries (with a separation distance $r_s > $ 100 AU) through close 
stellar encounters decreases the primordial binary fraction in dense clusters 
\citep{kroupa:95,kroupa:99} or (b) the influence of the temperature of the molecular 
cores on the fragmentation mechanism results in higher primordial binary fractions 
\cite{durisen:94,sterzik:03} in less dense clusters.  


Possibly connected with this problem is the question that is addressed here: why does
the binary frequency increase from $\sim$ 50\% for solar-type
field stars \cite{duquennoy:91,fischer:92} to $\sim$ 70\% for massive O- and B-stars 
\cite{abt:90,mason:98})?
%
%
In this paper we consider the possible reasons for the higher binary frequency of massive stars in 
comparison to intermediate mass stars. In principle there are only two possibilities:
either massive stars are more likely to be binaries to start with, or their binarity increases
within the first Myr of their existence in a significant way. 
In other words, the first possibility would mean that the star formation process is mass-dependent.
The question whether this is so is very difficult to answer, since there are still problems explaining 
how even a single star more massive than $\sim$ 10 \Msun\ can form \citep[for a review see][]{yorke:07}.
Here we restrict ourselves to the second possibility, i.e. we start with the assumption that the 
primordial  number of binaries is the same for solar-mass and massive stars and ask whether dynamical 
processes can lead to a sufficient amount of additional binaries to explain the difference in 
observed binary frequency between solar-mass and massive stars. 

To do this we revisit capture processes as a potential candidate for part of the binary formation.
In the early 90s, stellar dynamics simulations  \citep{clarke:91} of the ONC showed that dynamical 
interactions in a cluster cannot form a significant enough number of binaries from an initially 
single-star population to explain a 50\% or more binary population.
Indeed, we also find that for solar-type stars, capture processes rarely lead to bound systems.
However, it will be demonstrated here that the situation is different for {\em massive} stars. 
The combination of the higher capture cross section of the massive stars with the fact that they 
are predominantly located in the high density central regions of the cluster leads to the formation 
of massive binaries in addition to the primordial existing ones by capture processes in the early 
cluster development. 

As a model cluster we chose the Orion Nebula Cluster, because it is one of the densest close-by clusters
in the Galaxy, so if capture processes play any role one should find indications for it here.
In addition, it is observationally one of the best investigated clusters, so the detailed knowledge 
of its properties helps to limit the simulation parameters. What is known about the binary statistics in the 
ONC?        
In Orion the binary rate for solar-type stars is $\sim$ 50\% whereas the binary rate
for massive stars is $\sim$ 75 \% \cite{preibisch:99,koehler:06}.  The latter value has a high 
uncertainty upwards, as observations of the binary frequency of high-mass stars are usually obtained
from small-sized samples and inhomogeneity of the sample and/or selection effects in general cannot
be completely excluded. 
Correction for completeness, as done for lower mass stars, is difficult as the necessary assumptions of 
the binary mass ratio and orbital period distribution cannot be made.  This means that the observations 
could be equally well be interpreted as a 100\% binary rate for massive stars in the ONC \cite{preibisch:99,koehler:06}.
%
Possibly there exists a trend of 
decreasing multiplicity from the centre of the ONC outwards \cite{preibisch:99,koehler:06}.
One concludes that differences in the formation mechanisms of high-mass and low-mass stars 
multiple systems must exist. 

Although capture usually is no longer considered,   Moeckel \& Bally (2007a) studying single isolated 
encounter events of disc-surrounded stars only, suggested that 
encounter-induced capture could be a common event for  massive stars.  They estimated 
that the density in the ONC would suffice for capture processes alone to lead to 50\% binaries. 
Although we do not include the effect of discs in the simulations presented here, we perform 
the first systematic study of the stellar dynamics in an ONC-like system with the emphasis on the  
massive stars. It reveals that {\em massive stars frequently form ``transient bound systems''} (TBS) 
through gravitational interaction. Single such systems are indistinguishable from long-lasting classical 
binaries in observations, but in Section 4 we will see that some binary statistics results could be 
interpreted as effect of a population of capture-formed massive binaries.


What kind of binaries do form? Binaries can be sub-divided into three dynamical groups according to
the ratio between $E_b$, the binding energy of the binary, and  $E_{kin}$, the kinetic energy 
of the encounter with another star of the system: (i)
the wide, or soft, binaries with $E_b/E_{kin} <$ 1, (ii) the dynamically active binaries
$E_b/E_{kin} \sim$ 1, and
(iii) the tight or hard binaries $E_b/E_{kin} >$ 1. 
Soft and hard binaries are relatively well understood \cite[see][for a review]{heggie:03} and
\cite{heggie:75} and \cite{hill:75} previously summarized the dynamics of these systems with the
observation that ``soft binaries soften and hard binaries harden''.
Of these groups the active binaries with intermediate binding energies are least well understood. 
Such binaries couple efficiently to the cluster, and exchange energy with it. We will see that
it is this type of binary that the massive stars in the ONC primarily form in the
initial phases of the cluster development.
These active binaries or TBS are quite proficient in exchanging partners. 
%

Section 2 describes the numerical method applied. In Section 3 we determine the number of TBS formed by 
capture in the ONC followed by a detailed analysis of their properties. 
In Section 4 it is discussed how these findings might change the interpretation of the origin of a 
higher binary rate for massive stars.

\section{Method}

\begin{figure}
\resizebox{\hsize}{!}{\includegraphics[angle=-90]{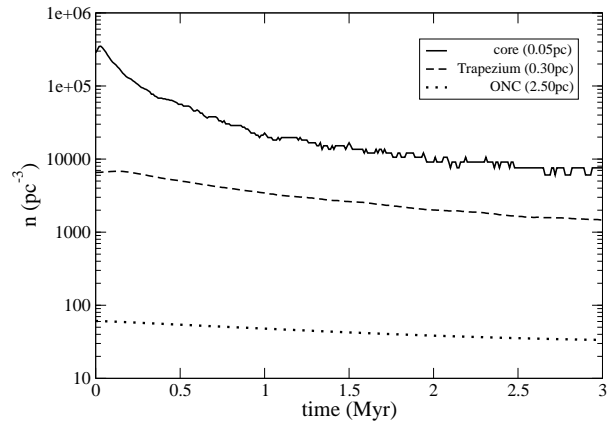}}
\caption{The development of the particle density as a function of time
for the central core, the Trapezium region and the entire ONC.}
\label{fig:density}
\end{figure}

For simplicity we start with a system initially consisting only of single stars - the influence 
of both primordial binaries and discs around the stars are excluded from this first study. Primordial 
binaries would make it more difficult to track the effect of the capture processes 
themselves, we will discuss in Section 4 in how the assumption of no primordial binaries
effects our results. The influence of discs can only be included if detailed knowledge of 
star-disc-star-disc interactions leading to binaries exists for the entire parameter space required. 
To our knowledge only single cases of such interactions have been modelled so far.

\begin{figure}
\resizebox{\hsize}{!}{\includegraphics[angle=0]{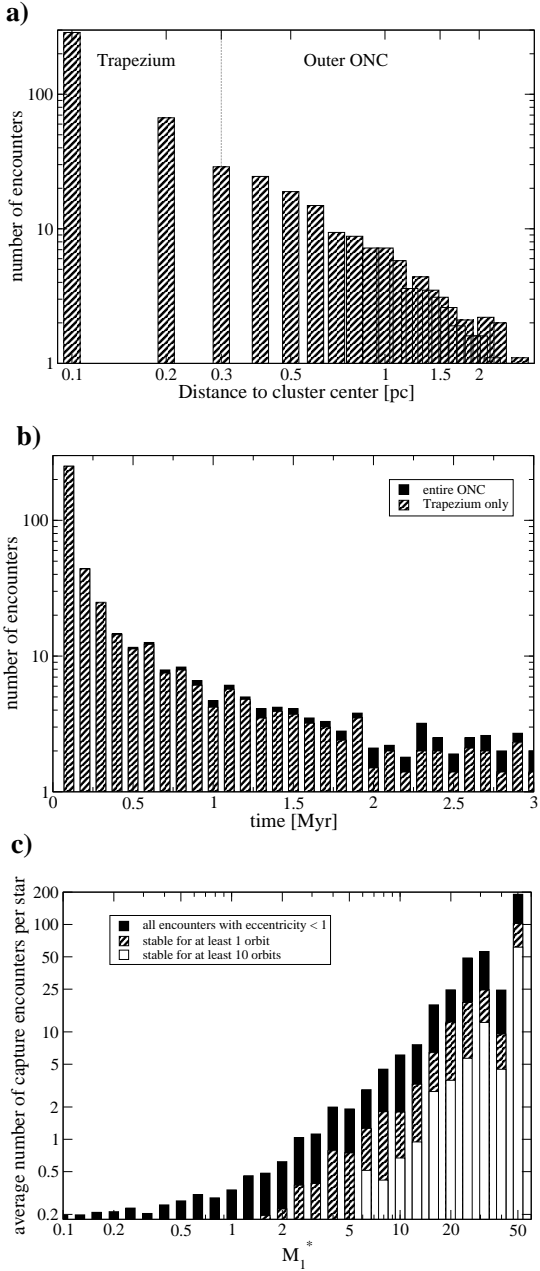}}
\caption{Number of encounters with $\epsilon <$ 1 as function of a) radial distance to cluster centre, 
b) time and c) primary mass M$_1^*$. In b) the number of encounters in the Trapezium and the entire ONC are shown.
In c) all encounters with  $\epsilon <$ 1 independent of the duration of the bound state are shown, but although 
the ones that keep at least bound for one (dashed line) and ten (dotted line) orbits.}
\label{fig:enc_mass}
\end{figure}

In the cluster simulations we followed the dynamical development of $\sim$ 4000 stars in an 
virial equilibrium situation, i.e. $Q_{vir}$=0.5, with a spherical density distribution 
$\rho(r) \sim r^{-2}$ using \nbodypp \cite{aarseth:book03,spurzem:mnras02}. 
Gas components and the potential of the background molecular cloud OMC~1 were neglected in these
simulations. The most massive 
star was assigned a mass $M^*$= 50 \Msun\ and all other stars' masses according to the mass 
distribution given by \cite{kroupa:01},
\be
\xi(M^*)= \left\{
\begin{array}{lrll}
0.035 M^*\,^{-1.3} & \mbox{ if } & 0.08&\le M^*\,<0.5, \\
0.019 M^*\,^{-2.2} & \mbox{ if } & 0.5&\le M^*\,<1.0, \\
0.019 M^*\,^{-2.7} & \mbox{ if } & 1.0&\le M^*\,<\infty. 
\end{array} \right.
\label{eq:imf}
\ee
The simulations were performed as described in \cite{olczak:06}. 
There the quality of the dynamical 
models was judged by comparing them to observational data at 1-2\,Myr, marking the range of the mean 
ONC age. The quantities of interest were: number of stars, half-mass radius, number densities, 
velocity dispersion and projected density profile. 
In the initial cluster the central number density within 0.05 pc was 2.85 $\times$ 10$^5$ pc$^{-3}$ and
the velocity dispersion 2.4 km/s. There is a fast development of the number density (see Fig.~\ref{fig:density})
which peaks in the core at $\sim$ 0.03Myr.

However, we found it necessary to improve the tracking of the cluster centre for comparison
to observational data. Previously we used the centres-of-mass to define the cluster centre. However, as 
the simulations contain only a limited number of stars,
``escapers'' kicked out of the cluster due to strong interactions, can temporarily shift the 
centre-of-mass before they are removed at large distances.
Using the (mass or number) density centre is a more reliable alternative, especially 
since observers usually use either the brightness concentration or the maximum projected number 
density of objects as a reference point.
However, 
since the local particle density changes strongly with time, so does the estimated 
density centre, and is additionally influenced by the choice of the number of 
nearest neighbours to some degree. 

Therefore we use the density centre but introduce the following smoothing algorithm:
Instead of using all simulation particles to define the centre of mass, particles at large 
distances from the cluster centre are excluded. On the other hand this sub-sample has to be large enough 
that proper statistics are possible and that the most massive stars are not too dominant. This is achieved
in practice by requiring that the sub-sample contain at least 10\% of the total stellar population 
and that the most massive star must represent at most 10\% of the mass of this sub-sample. 
The centre-of-mass of the sub-sample serves as the new cluster centre.
This algorithm combines the advantages of keeping the cluster centre close to 
the density centre and taking a larger stellar sample into account to smooth the strong temporal
fluctuations.

Having defined the cluster centre in such a way, simulations have been selected as valid representations 
of the ONC if they fulfilled the following criteria:
\begin{itemize}
\item{The projected density distribution at 1\,Myr has to match the data from 
Hillenbrand \& Hartmann (1998) and McCaughrean et al. (2002) within the statistical errors.}
\item{The most massive star, \COri, must be located inside the Trapezium Cluster ($R_{\small\textrm{TC}}=0.3$\,pc) for at least 1\,Myr, the estimate of the mean age of the ONC.}
\end{itemize}

Altogether we have selected 40 simulations according to the above criteria.
During the course of these simulations we track all encounters with eccentricity $\epsilon <$ 1, 
noting the stellar mass ratio, periastron, eccentricity and duration of the bound state. It is not checked 
for bound systems with more than two partners, as we want to concentrate on the binary rather than
the general multiplicity rate in this investigation.

Despite the large number of simulations and the resulting several hundred TBS formed per simulation run, 
one still has to deal with small-number statistics here. There are two reasons for this:  the parameter space 
in mass ratio, periastron, etc. is very large and it is often necessary to restrict the sample to a certain 
time interval or for systems to stay stable for a certain duration. This leads to the difficulty that
the results in the following section are sometimes presented for different time-intervals.
We only present results where we have at least 20 events per bin to guarantee statistical significance. One
should keep in mind that observations usually have much fewer objects to work with when determining binary properties.

\section{Results}

\begin{figure}
\resizebox{\hsize}{!}{\includegraphics[angle=0]{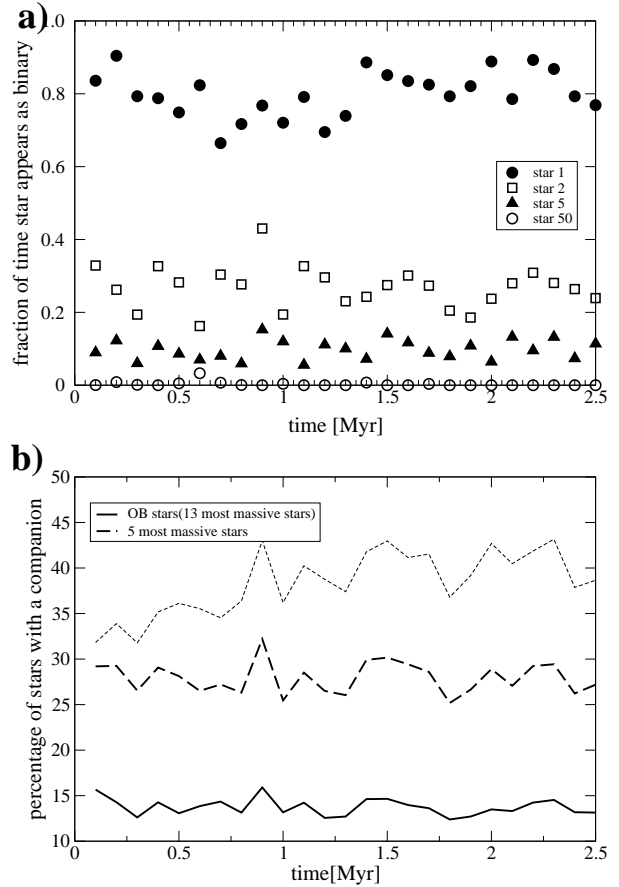}}
\caption{a) shows the fraction of time a TBS would appear as a binary for different
primary stars. The distribution of the stellar masses is not fixed but varies. This done to give a
better representation of the IMF on average. Therefore no fixed stellar masses can be given here, 
but the higher star number is equivalent to lower stellar mass. In b) the percentages of the 
5 (dashed line) and 13 (full line) most massive stars are shown that 
are part of an encounter-induced binary in a time interval of 0.1 Myrs as a function of cluster age.
The thin dashed line shows the case where these massive TBS are considered twice.}
\label{fig:binarity}
\end{figure}

All encounter events with an eccentricity $\epsilon <$ 1 --- hitherto called ``capturing encounters'' ---
are recorded. It turns out that these capturing encounters mostly happen early on in the cluster 
development, close to the cluster centre and mainly involve one of the most massive stars 
(see Fig.~\ref{fig:enc_mass}a-c).
This is exactly what one expects: close to the cluster centre the stellar density
is highest, even more so early on when the cluster initially contracts (see Fig.~\ref{fig:density}) 
and the most massive stars 
dominate this area gravitationally. Specifically,  during the first Myr there are on 
average more than 350 capturing encounters in the Trapezium region, in contrast to 
$\sim$ 150 in the rest of the ONC, which covers a thousand times larger spatial volume (see 
Fig.~\ref{fig:enc_mass}a). During the first 0.3 Myr of the cluster development more
than 300 capturing encounters occur; afterwards less than 15 capturing events happen
in any 0.1 Myr time interval (see Fig.~\ref{fig:enc_mass}b).
Whereas the most massive star has on average of the order of 200 capturing 
encounters in the first 5 Myr, a star with $M_1^*= 1$\Msun\ has less than one. In addition, a higher 
percentage of the capturing encounters involving the massive stars lead to relatively stable 
configurations. If the primary is of mass  $M_1^*$=50\Msun, about half of the forming systems 
stay stable for at least one orbit and a quarter for at least 10 orbits. By contrast, for a primary of 
mass $M_1^*$= 1\Msun\ about a third stay stable for one orbit and less than 10\% for more 
than 10 orbits.  Here and in the following the word ``primary'' is used for the more massive rather 
than the more luminous star, in contrast to convention in observations.     

We are now in a position to ask which binarity would be observed for clusters like our ONC model
cluster. Fig.~\ref{fig:binarity}a) shows the probability of finding the most massive stars in a 
bound state as a result of a capturing process. It can be seen that the most massive star would 
in the majority of cases be found as a partner in a 
bound state. The second most massive star would be likely to be bound in $\sim$ 30$\%$ of all 
cases. For stars with lower mass the likelihood to be found in such a capture-induced bound state
decreases rapidly with decreasing mass (see star 5 and 50 in Fig.~\ref{fig:binarity}a, where higher numbers 
are equivalent to lower masses). 

Fig.~\ref{fig:binarity}b shows the likelihood of the 5 and 13 (number of OB stars in the ONC) 
most massive stars of the cluster to have at least one companion (here the  percentage is shown) 
as a function of time. 
At 2 Myrs 10-15$\%$ of the all OB stars and 25-30$\%$ of the five most massive stars would form
TBSs due to the interaction dynamics in the cluster and would appearing as 
binaries. Here a TBS is taken into account only once if both stars were massive.
The thin dashed line in Fig.~\ref{fig:binarity}b shows the case where these massive TBS are
considered twice. This is equivalent to the likelihood of a specific star to be in a TBS.
For the five most massive stars this likelihood to be in a TBS is $\sim$ 40 $\%$ at 2 Myr.
Fig.~\ref{fig:binarity}b agrees with the result by Moeckel \& Bally (2007b) who found 
an expected binary fraction of 12\% for a star with $M_*=20\Msun$ at 0.5 Myr.
So for the most massive stars of the cluster there is a high likelihood
of it being a TBS and appearing as a binary. The number of TBS is actually sufficient to explain the
difference in binary rates between massive and solar-mass stars. But what are the properties of these 
capture-formed TBSs?

\begin{figure}
\resizebox{\hsize}{!}{\includegraphics[angle=0]{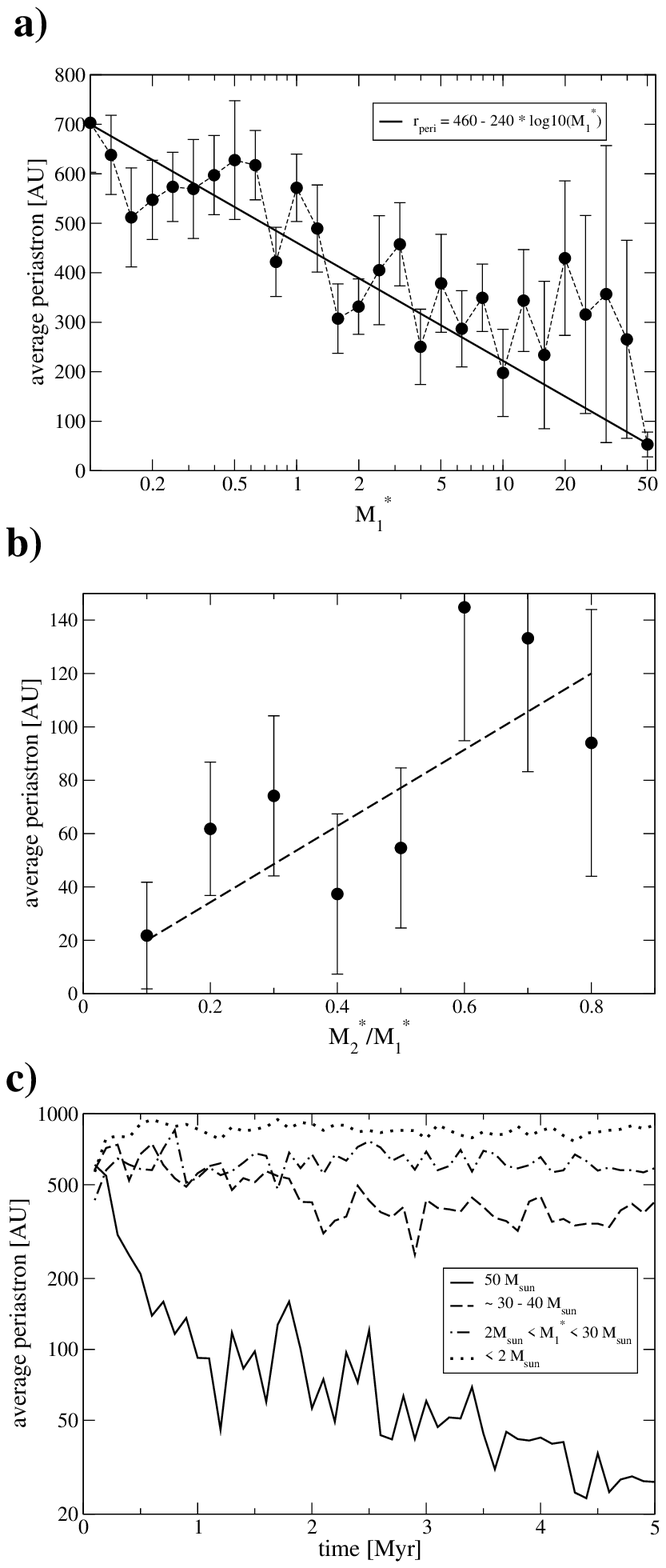}}
\caption{Average periastron as a function of a) the primary mass $M_1^*$, b) the mass ratio $q = M_2^*/M_1^*$
and c) time. Here only encounters where considered were the binary remained bound for at least one entire orbit. 
These are averaged over all times up to 2Myr. In b) the periastra of the most massive star for the time interval 
0.3-3.0 Myr are shown as an exemplary case. c) shows the average periastron for different primaries separately: 
$M^*_1$=50 \Msun, 30 \Msun $ < M^*_1 <$ 40 \Msun, 2 \Msun $< M^*_1 <$ 30 \Msun, and $M^*_1 <$ 2 \Msun.}
\label{fig:peri_mass}
\end{figure}

We start with the periastra of the transient bound systems.
Fig.\ref{fig:peri_mass}a) shows the average periastron as a function of the primary mass $M_1^*$. 
The values are averaged over the first 5 Myr of the cluster development. 
It can be seen that the average periastron declines with increasing primary mass from $\sim$ 
700 AU for stars with $M_1^* < $  0.1\Msun\ to $\sim$ 60 AU for $M_1^* $=50\Msun, reflecting the fact 
that massive stars form tighter bound systems more easily. 
The decline in periastron 
can be approximated by a linear dependence on log10($M_1^*$).

\begin{figure}
\resizebox{\hsize}{!}{\includegraphics[angle=0]{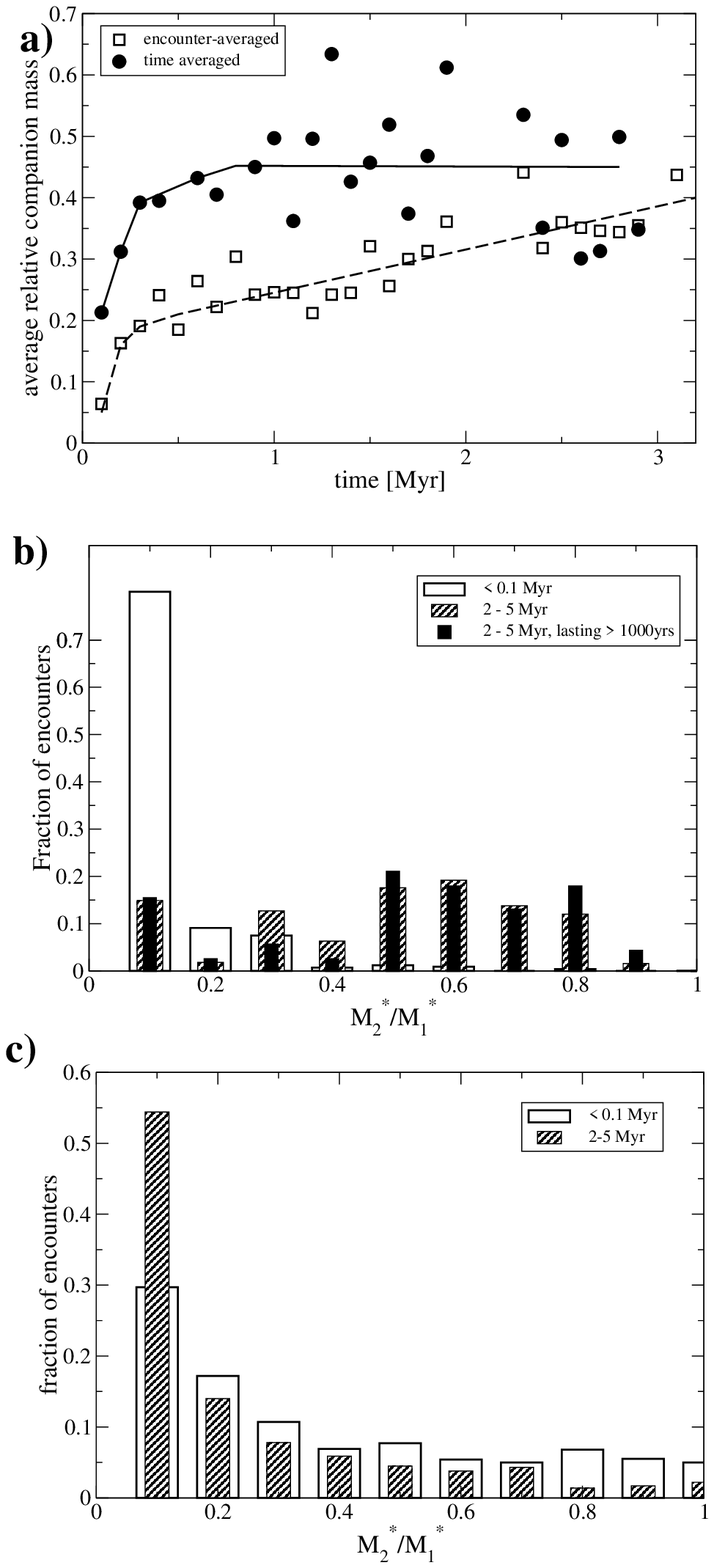}}
\caption{a)The average relative companion mass $M_2^*/M_1^*$ as a function of time for 
capturing encounters for $M_1^*$ = 50 \Msun. The line is no proper fit, but just meant as a guide line. b) Distribution of companion mass up to 
0.1 Myr and 2 - 5 Myr for the most massive star and c) for stars with $M_1^* <$ 10 \Msun.}
\label{fig:m2_dist}
\end{figure}

Fig~\ref{fig:peri_mass}b) shows the average periastron as a function of the mass ratio $q = M_2^*/M_1^*$ 
for the most massive star for the time interval 0.3-3.0 Myr. Due to the random selection process for the 
stellar masses in the initial
cluster set-up, the second most massive star usually has a mass in the range of 30-40 \Msun, so the 
statistics become increasingly worse for  q $>$ 0.7. To a good approximation the average periastron 
increases linearly with higher mass ratios from $\sim$ 50-65 AU for $q$ = 0.1 to $\sim$ 160 AU for 
$q$ = 0.7. This reflects the fact that lower mass stars can only replace a more massive star if they orbit on a 
smaller semi-major axis. In other words, the more equal the masses of the stars bound to each other, 
the larger the periastron of a relatively 
stable configuration can be.

The average periastron as a function of cluster age is shown in 
Fig.~\ref{fig:peri_mass}c). Generally there is a wide spread in periastra data obtained in single simulation runs.
TBS with large periastra are usually of short duration, so it seems more appropriate to average over
the periastron weighted by the duration of the bound state. This reflects the 
observations which are more likely to detect long-lasting bound systems than short 
ones. The average periastron of the 50 \Msun-star strongly depends on the cluster age, decreasing from $\sim$ 
500-600 AU at 0.1 Myr to $\sim$ 20-30 AU at 5 Myr. Exchange processes lead to increasingly stronger bound TBS
over time. For the stars in the mass group $M_1^*$ $\sim$ 30-40 \Msun\  the trend is less 
pronounced, decreasing from  $\sim$ 700 AU at 0.1 Myr to $\sim$ 350-450 AU at 5 Myr. For lower mass stars 
the periastron does not significantly alter over the considered timespan and is the larger for lower 
primary masses. At 1-2 Myrs, the most likely age of the ONC, the average periastron of the most 
massive TBS is between 50-200 AU.

Next we consider the two properties that observations usually come provide: the
average mass ratio $q=M_2^*/M_1^*$ and the binary frequency.

Fig.~\ref{fig:m2_dist}a shows the average relative mass ratio $M_2^*/M_1^*$ of the TBS for the most massive stars. 
It can be seen that in the encounter-averaged case the 
relative mass of the companion increases from very low values ($< 0.1 M_1^*$) in the early cluster development 
to $q$-values $\sim$ 0.4 at 3 Myr. If we weigh the q-values by the encounter duration, the general trend
remains the same but shifts to somewhat higher values.
Our simulations show in addition the difference in this development for primaries with low and high mass.
Looking at the distribution of companion masses one sees in Fig.~\ref{fig:m2_dist}b  that
for the most massive star at early times (t$<$0.2Myr) mass ratios $q <$0.1 dominate, 
whereas capturing encounters with more massive stars become increasingly important later on in the
cluster development.
The mass ratio in these TBS develops from an initial preference of low-mass 
companions to companions with high mass. At an cluster age of 2-5Myr the maximum of $q$ lies in the
range of 0.6-0.8, which is even more pronounced, if we consider only TBS which last at least 1000 yrs.
The value of $q$ is not the expected equal mass binary value of 1, because the mass of the second most massive
star varies in the different simulations between 30-40\Msun . In addition, the most massive star usually
captures just one of the 10 most massive stars as companion and not necessarily the second most massive star.  

This result of high q-values later on in the cluster development correlates very well with the observations 
which show that for the massive stars the mass ratio $q$ seems to be much smaller in young cluster than for 
field stars \cite{goodwin:06}. How quickly this development happens will depend crucially on the cluster properties 
such as its density.

\begin{figure}
\resizebox{\hsize}{!}{\includegraphics[angle=-90]{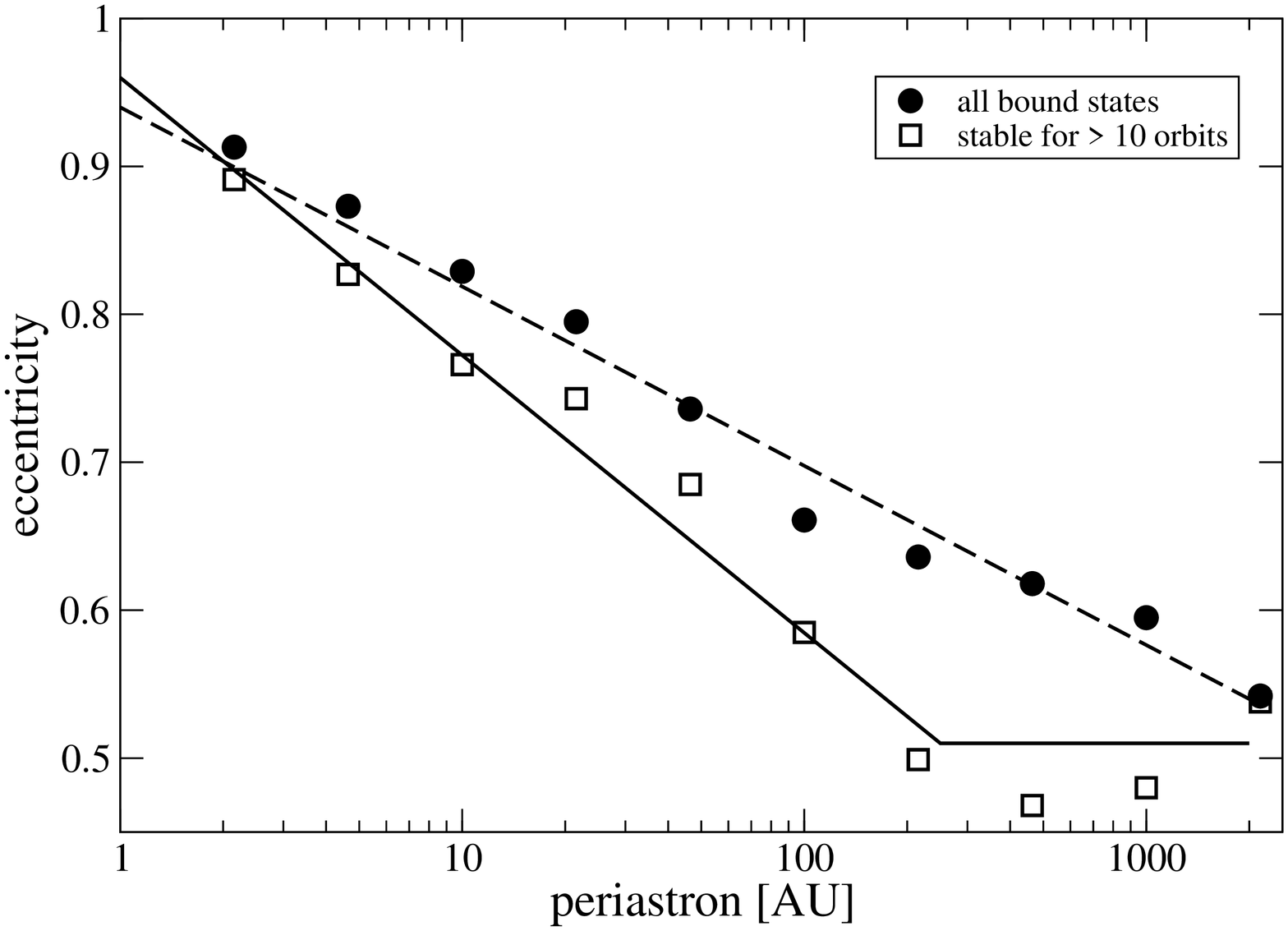}}
\caption{Eccentricity as a function of periastron. Here only encounters where considered were the
binary remained bound for at least one entire orbit. These are averaged over all times up to 2Myr. Only
 $M_2^*/M_1^*<1$ are considered. The line is no proper fit, but just meant as a guide line.}
\label{fig:ecc_peri}
\end{figure}

What about the eccentricity in these bound systems?
Fig.~\ref{fig:ecc_peri} shows a very clear connection between eccentricity and periastron:
the larger the periastron the smaller the average eccentricity. This is simply what one would expect
as a consequence of two-body relaxation.  Obviously, whereas a binary with a 
1 AU periastron can tolerate a 0.9 eccentricity, a binary with a 2000 AU periastron has on average 
only a 0.55 eccentricity in the here investigated system. If we consider only systems remaining bound 
for at least 10 orbits, the decline is even faster, but levels of at about 200 AU at an eccentricity of 
$\sim$0.5. This faster decline reflects that large periastron TBS only stay stable if their eccentricity is low, 
otherwise they are likely to undergo strong interactions with the other cluster stars. We saw before that the
TBS at 1-2Myrs have on average a periastron of 50-200AU. Using Fig.~\ref{fig:ecc_peri} this is coupled to
an average eccentricity $\epsilon \sim$ 0.5-0.6.

\begin{figure}
\resizebox{\hsize}{!}{\includegraphics[angle=-90]{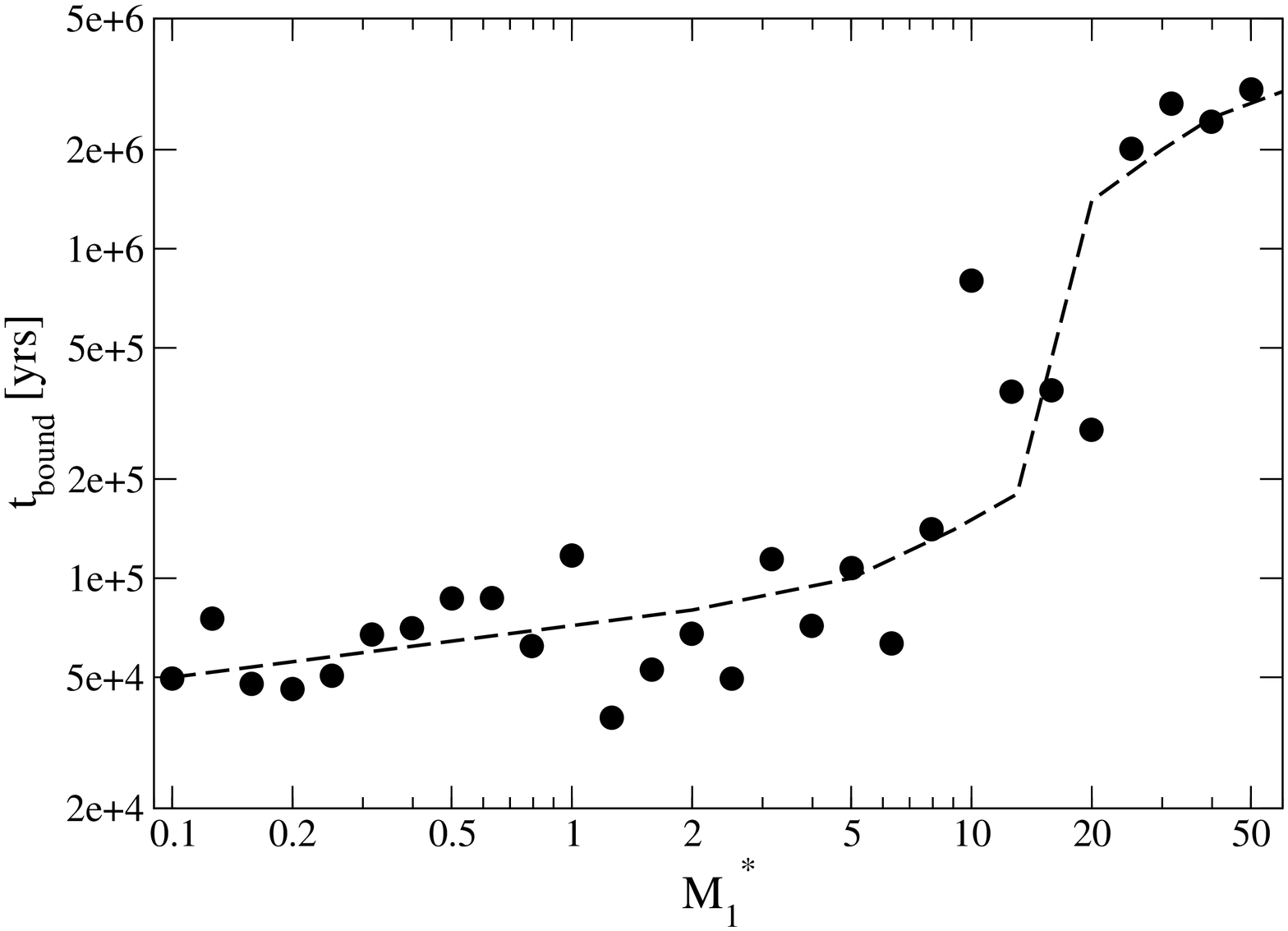}}
\caption{Average duration of a bound state, $t_{bound}$, as function of the primary mass $M_1^*$.
The line is no proper fit, but just meant as a guide line.}
\label{fig:binary_time}
\end{figure}

The trend towards more stable configurations, especially for the most massive stars, is most
dramatically reflected in the average duration of TBS. 
%
%
The dependence of the average duration of a bound state on the stellar mass $M_1^*$ of the primary can be
seen in Fig.~\ref{fig:binary_time}. It shows that 
generally the bound states last on average  longer the higher the primary mass. This is most pronounced for stars with
$M_1^* > $ 10 - 20 \Msun\ where the bound state lasts on average several times $10^6$yrs in contrast to
bound state durations of on average $< 10^5$ yrs for lower mass stars. So apart from the very early
stages of the cluster developement ($<$ 0.1Myr) these systems stay stable for many orbital periods.

The duration of the binary phase $t_{bound}$ increases for larger  
$M_2^*/M_1^*$,  these are the more stable configurations that are formed later in the cluster development.
For the same reason, we see in our simulations an increase of $t_{bound}$
with larger periastra and smaller eccentricities (a strong increase for $\epsilon < 0.2$).  
This formation of more stable configurations is reflected 
by a dramatic increase of the average duration of a bound state with cluster age. Whereas bound states last on average 
only a few 10-100 yrs at $t_{cluster}$ = 0.1 Myr, it rises to a few $10^6$ yrs at  $t_{cluster}$ = 2 Myr.
So in very young clusters these massive stars would {\em appear} as binaries but are actually just running 
through a succession of TBS. 

As the duration of the bound state is a strong function of time so is the number of orbits. 
It is initially higher but increases less rapidly for lower mass primary stars. So that after 1 Myr the binaries 
involving the most massive stars stay bound for a larger 
number of orbits ($\sim$ several 1000).

\section{Discussion and Conclusions}

The results of our simulations of capturing processes in an ONC-like system can be summarized by the following:
In cluster environments similar to the ONC, massive stars have a much higher probability of 
involvement in a capturing encounter than solar-mass stars. Assuming a cluster age of 1-2 Myr,
at least 10-15\% of the OB stars in the ONC are in a bound state caused by capturing processes, 
with the average periastron of the so-formed TBS being $r_p \sim$ 50-200 AU, 
the mass ratio $q \sim$ 0.4-0.5 and the eccentricity $\epsilon \sim$ 0.6.
In observations these bound systems would be regarded as normal binaries, even if the bound state is 
of short duration($\sim 10^{3}$ - 10$^6$ yr). 

So we have to discuss two questions: i) is the number of TBS formed sufficiently high to explain the difference 
in binary between solar-type and massive stars and ii) are the observed properties of these TBS in accordance 
with observations?

For simplicity the simulations presented here started with all stars initially being single. In reality, 
there exists a primordial binary fraction. We saw that for massive stars, capture processes quickly lead 
to additional bound systems, so that  there are two binary 
populations present - the primordial binaries and the TBS formed by capture processes. 
As our simulations show that at least 10-15\% of the OB stars in the ONC are in a bound state caused by capturing 
processes, the observed higher binary frequency for massive stars compared to solar-mass stars
must reflect therefore, at least partly, the presence of these transient capture-formed binaries. 

The value of 10-15\% can be regarded as a lower limit, because most, if not all, stars in the cluster 
are initially surrounded by discs. 
The large radii of the discs compared to the stellar size results in a much larger cross-section 
for encounters, an effect that has not been included here. The larger capture cross-section will 
influence the overall dynamics, leading to more bound systems. 
Moeckel \& Bally (2007a,b) studied the disk-assisted capture formation of binaries for stars of $M_1^*$ = 20 \Msun\ 
and estimated that in the Trapezium region, capture could account for $\sim 50 \%$ of massive stars 
having a companion after 1 Myr. 
Since the presence of discs around the stars increases the capture cross-section, it is even possible 
that the fraction of primordial binaries is the same for solar-mass 
stars as for massive stars and that the observed higher binarity could be {\em exclusively} due to 
the encounter-induced capture processes. To clarify this, future modelling of the cluster dynamics should 
include the effect of discs by some means.

However, it is most likely that discs only influence capture processes significantly in the very early 
cluster development, because observational examples for discs around massive stars are usually only found for young, 
still embedded, protostars \cite{cesaroni:99,tafoya:04,patel:05,kraus:07}. Massive stars in a more 
evolved stage, like in the Trapezium, are usually not surrounded by discs. 
Processes like photoevaporation \cite{clarke:07} and gravitational interaction \cite{pfalzner:aa06}
are responsible for the disc destruction. Observational and theoretical evidence point to
lifetimes of discs around massive stars of approximately $\sim$ 10$^5$ - 10$^6$ years. 

The second question whether the observed properties of these TBS are in accordance 
with observations is difficult to answer. The reason is that the ONC contains only 15 OB-stars
and as we have shown, the observed binaries are a combination of primordial binaries and
those formed in capture processes. We would expect here  2-4 of the massive stars
to be part of a TBS - clearly a number where statistics is impossible.
Another problem is that observing the Trapezium the surface density of stars is such
that the visual binary population at periastra of a few 100 AU is difficult to distinguish 
from unassociated stars with current spectroscopic methods. However, future interferometric
observations of proper motions - with, for example the VLTI, - should be able to provide the
neccessary data.  

Interesting in this context is that recent observations \cite{kraus_s:07} indicate that \COri\ itself is a binary 
with a massive companion, possibly with \COri\ having a mass of 34 \Msun\ and the companion mass being 15.5 \Msun. 
Naturally one cannot decide whether a specific binary has formed primordially or as result of a
capture process. In this specific case the high mass ratio would speak for a capture process, but the low
periastron of $\sim$1.5 AU could occur in both cases  - capture-formed or primordial binaries.

So although the ONC has the advantage that many observational results are available, it has the
disadvantage that it does not contain a high number of massive stars. 
However, statistical properties in other cluster --- especially in very young ones --- could 
provide signatures of capture processes. Properties where one would expect observable differences between
primordial and capture-formed binaries are the periastron and the mass 
ratio distribution. Our simulations show that the periastra of the encounter-induced bound 
systems are usually in the range of a few 10 - 2000 AU for the first Myrs of evolution, so they clearly belong 
to the so-called wide binaries. 
The observed over-abundance of wide binaries among PMS stars compared to field stars \cite{patience:02} could 
possibly be a tracer for these capture-formed binaries. As we have shown here the average periastra in TBS with a massive
primary become smaller ($<$ 100 AU) for older clusters, so one should be able to trace 
this effect by observing the binary frequency of massive stars in clusters of different age but similar structure.
The over-abundance of wide binaries should shift to smaller periastron values and become less pronounced with 
increasing age. 
The latter is due to the shorter lifetime of the massive stars, so these binaries would naturally disappear
if the cluster is beyond a certain age.  

Our simulations indicate that capture processes lead to an increase in the mass ratio with cluster age.
Investigations of the mass ratio in binaries with massive primary to date give a complex picture: while small $q$
(consistent with random pairing) are observed in the ONC \cite{preibisch:99}, 
in rich clusters most O stars are believed to exist as short-period binaries with $q \simeq 1$
\cite{garcia:02}. Mass-ratio distributions not consistent with random pairing were also reported
by Kouwenhoven (2005) for A and late-type B binaries in Scorpius OB2.

So ideally in future this kind of 
investigation should be performed for a cluster containing a high number of O stars.
The question is: what can one expect in systems other than the ONC? In lower density and 
open clusters the number of bound systems formed by capturing will be too insignificant to contribute 
to the total binary fraction. However, in high-density clusters like the Arches cluster near 
the Galactic centre the central mass density ($\sim$ 3 $\times$ 10$^5$ \Msun pc$^{-3}$ 
(Figer et al. 2002)) is about ten times higher than in the 
Trapezium region, so capturing encounters could play an important role. However, this is
difficult to predict because a number of other cluster properties influence the capture rate. Apart from 
the cluster density, the velocity dispersion and the degree of mass segregation will be significant
factors. In the Arches cluster  the velocity dispersion is not very well known,
only an upper limit of 22 km s$^{-1}$ exists. If the real velocity dispersion is close to this
upper limit, it might mean a large number of high velocity encounters, which would decrease
the capture rate. The relative importance of these different cluster properties for the capture
rate need further investigation. 

Since the primordial binary fraction seems to depend strongly on the cluster 
environment, it is important to compare relative values. One way to find indications for the 
capture-formed massive binaries would be to compare the relative binary frequency of solar-type and massive 
stars in clusters of different stellar density. If capturing is an important process, the higher 
frequency of capturing encounters in dense systems would lead to a higher fraction of massive binaries 
to solar-type binaries there. However, one has to be careful with very dense systems containing many OB-stars, 
since higher velocity dispersion could counteract this process.

The study performed here can only be regarded as a first step towards understanding the
significance of capture processes in young dense clusters. In future some points need 
further investigation. First, we concentrated here on the closest bound partner, whereas observations as well
as our simulations indicate that multiple systems are a common feature for massive stars.
Second, simulations should start with a finite primordial binary fraction and see how it develops.
The massive stars would still capture partners, but it is also known that
the dynamical evolution of dense clusters has a destructive effect on primordial binaries 
\cite{kroupa:95,kroupa:99,kroupa:01}. It would be important to see which of these competing effects dominates. 
Third, as pointed out above in order to improve the statistics, simulation of systems containing 
a higher number of OB stars are required.
Finally, disc effects have to be included as despite their short lifetimes around massive stars 
they could not only increase the capture-induced binary rate but also influence 
the periastron and eccentricity of the formed systems. From our results it is apparent that it is not uncommon for
the periastron in capture-formed bound systems to be just a few 10 AU. This means that the secondary 
star would pass through the disc of the primary.  Repeated disc passages can reduce the periastron and 
eccentricity through gravitational drag. For our
simulations this would mean an even faster development in the initial stages.


\section*{Acknowledgments}

We are grateful to R.Spurzem for providing the Nbody6++ code for the cluster 
simulations and want to thank H.Zinnecker for his very helpful comments.
Simulations were partly performed at the John von 
Neumann Institute for Computing, Research Centre J\"ulich, Project HKU14.

\bibliographystyle{apj}

\end{document}